\documentclass[aps,prb,preprint,superscriptaddress,showpacs,preprintnumbers,amsmath,amssymb]{revtex4}

\usepackage{graphicx}
\usepackage{bm}

\begin{document}

\title{Magnetic structure and critical behavior of GdRhIn$_{5}$:
resonant x-ray diffraction and renormalization group analysis}

\author{E. Granado}
\email{egranado@ifi.unicamp.br}
\affiliation{Instituto de F\'{i}sica ``Gleb Wataghin,'' UNICAMP,
13083-970, Campinas-SP, Brazil}
\affiliation{Laborat\'{o}rio Nacional de
Luz S\'{i}ncrotron, Caixa Postal 6192, CEP 13084-971, Campinas,
SP, Brazil}

\author{B. Uchoa}
\affiliation{Laborat\'{o}rio Nacional de
Luz S\'{i}ncrotron, Caixa Postal 6192, CEP 13084-971, Campinas,
SP, Brazil}

\author{A. Malachias}
\affiliation{Laborat\'{o}rio Nacional de
Luz S\'{i}ncrotron, Caixa Postal 6192, CEP 13084-971, Campinas,
SP, Brazil}

\author{R. Lora-Serrano}
\affiliation{Instituto de F\'{i}sica ``Gleb Wataghin,'' UNICAMP,
13083-970, Campinas-SP, Brazil}

\author{P. G. Pagliuso}
\affiliation{Instituto de F\'{i}sica ``Gleb Wataghin,'' UNICAMP,
13083-970, Campinas-SP, Brazil}

\author{H. Westfahl Jr.}
\affiliation{Laborat\'{o}rio Nacional de
Luz S\'{i}ncrotron, Caixa Postal 6192, CEP 13084-971, Campinas,
SP, Brazil}

\begin{abstract}

The magnetic structure and fluctuations of tetragonal GdRhIn$_{5}$ were studied by resonant x-ray diffraction at the Gd $L_{II}$ and
$L_{III}$ edges, followed by a renormalization group analysis for this and other related Gd-based compounds, namely Gd$_{2}$IrIn$_{8}$
and GdIn$_{3}$. These compounds are spin-only analogs of the isostructural Ce-based heavy-fermion superconductors. The ground state
of GdRhIn$_{5}$ shows a commensurate antiferromagnetic spin structure with propagation vector $\vec{\tau}=(0,\frac{1}{2},\frac{1}{2})$, corresponding to
a parallel spin alignment along the $\vec{a}$-direction and antiparallel alignment along $\vec{b}$ and $\vec{c}$. The spin
direction lies along $\vec{a}$. A comparison between this magnetic structure and those of other members of the 
$R_{m}$(Co,Rh,Ir)$_{n}$In$_{3m+2n}$ family ($R=$rare earth, $n=0,1;m=1,2$) indicates that, in general, $\vec{\tau}$ is
determined by a competition between first- ($J_{1}$) and second-neighbor ($J_{2}$) antiferromagnetic (AFM) interactions. While a large $J_{1}/J_{2}$
ratio favors an antiparallel alignment along the three directions (the so-called $G$-AFM structure), a smaller ratio favors
the magnetic structure of GdRhIn$_{5}$ ($C$-AFM). In particular, it is inferred that the heavy-fermion superconductor CeRhIn$_{5}$ is
in a frontier between these two ground states, which may explain its non-collinear spiral
magnetic structure. The critical behavior of GdRhIn$_{5}$ close to the paramagnetic transition at $T_{N}=39$ K was also studied
in detail. A typical second-order transition with the ordered magnetization critical parameter $\beta=0.35$
was experimentally found, and theoretically investigated by means of a renormalization group analysis.
Although
the Gd $4f^{7}$ electrons define a half-filled, spherically symmetrical shell, leading to a nearly isotropic spin system,
it is argued that a significant spin anisotropy
must be claimed to understand the second order of the paramagnetic transition of GdRhIn$_{5}$ and the related compound
Gd$_{2}$IrIn$_{8}$.

\end{abstract}

\pacs{75.25.+z, 75.40.Cx, 75.50.Ee, 61.10.Nz}

\maketitle

\section{INTRODUCTION}

The recent discovery of a new class of heavy-fermion superconductors, Ce$_{m}$(Co,Rh,Ir)$_{n}$In$_{3m+2n}$
($n=0,1;m=1,2$),\cite{Hegger,Petrovic1,Petrovic2,Pagliuso1,Thompson1} 
has triggered extensive research on their physical properties. While it is well established that
the spin fluctuations are key ingredients to determine the superconductivity
in this family,\cite{Moriya} the details of the pairing mechanism are not
presently known.\cite{Thompson2} Nonetheless, the relatively large critical temperatures found for
some compounds ($T_{c}=2.3$ K for tetragonal CeCoIn$_{5}$ is a record-high value for Ce-based
heavy fermion systems), and the evolution of this property with the ratio of $c$ and $a$ lattice parameters\cite{Pagliuso3,Bauer}
indicates that ingredients closely related with the crystalline environment must be taken into account.
This connection can be further explored by direct investigations of the sensitivity of the electronic structure on small deviations of the crystalline environment,
as well as of the influence of the such environment on the nature and magnitude of the spin structures
and the fluctuations that presumably mediate the superconductivity. In line with the second approach, the
magnetic structures below $T_{N}$ were resolved for a number of compounds, revealing interesting trends. The cubic
CeIn$_{3}$\cite{Benoit,Lawrence} and tetragonal Ce$_{2}$RhIn$_{8}$\cite{Bao2} show commensurate magnetic structures with antiferromagnetic (AFM) alignment of Ce spins
along the three nearest-neighbor directions (here called $G$-AFM phase in analogy to the nomenclature used in the manganites\cite{Wollan}),
and CeRhIn$_{5}$ forms an incommensurate spiral along the $c$-direction while still
keeping the AFM coupling in the $ab$-plane.\cite{Bao1,Curro} Recently, coexisting magnetic orders were observed in
the alloy system CeRh$_{1-x}$Ir$_{x}$In$_{5}$ in the interval $0.25 \lesssim x \lesssim 0.6$ where AFM
and superconducting phases overlap.\cite{Christianson} The first AFM phase is identical to the spiral phase of CeRhIn$_{5}$,
while the second one shows AFM alignment along the three nearest-neighbor directions such as in
CeIn$_{3}$\cite{Benoit,Lawrence} and Ce$_{2}$RhIn$_{8}$.\cite{Bao2} These results indicate that distinct AFM ground states
compete in this system, and that such competition may favor or be related with superconductivity.

Magnetic phenomena and their connection to crystal structures can be further explored by a thorough
investigation of other members of the $R_{m}$(Co,Rh,Ir)$_{n}$In$_{3m+2n}$ family ($R=$rare earth $\neq$ Ce, $n=0,1;m=1,2$).
Since such compounds are not heavy fermions and/or superconductors, the knowledge thus obtained may be taken as a starting point to understand 
the more complex and rich behavior of the Ce-based compounds.
Previous macroscopic studies on the $R_{m}$(Co,Rh,Ir)$_{n}$In$_{3m+2n}$ family ($R=$ Pr, Nd, Sm, Gd, and Tb) indicate
that the evolution of the N\'{e}el temperature ($T_{N}$) with $R$ does not follow the de Gennes scaling,\cite{Pagliuso,Serrano} suggesting
that crystal field or other anisotropy effects may be important to determine the critical temperatures and perhaps even the
magnetic ground state in this system. Neutron scattering studies on NdIn$_{3}$ indicate a ground state with AFM alignment
along two nearest-neighbor directions and FM alignment along the third direction, the so-called $C$-AFM structure.\cite{Mitsuda}
Additional phases with modulated moments along the FM directions have also been identified below the N\'{e}el temperature
for this compound.\cite{Mitsuda,Amara}
Similar $C$-AFM ground states have been identified in NdRhIn$_{5}$\cite{Chang}, Gd$_{2}$IrIn$_{8}$\cite{Granado} and
TbRhIn$_{5}$.\cite{Serrano2}

In continuation of the attempt to build a minimum comprehension of the magnetism of a spin-only
system ($L=0$) under similar crystal environments to Ce$_{m}$(Co,Rh,Ir)$_{n}$In$_{3m+2n}$ ($n=0,1;m=1,2$) heavy-fermion superconductors,
we carried out an experimental and theoretical investigation of the magnetic structure and fluctuations of
a GdRhIn$_{5}$ single crystal. We note that the magnetism of Gd-based compounds is expected to be particularly simple,
due to the absence of large orbital moments and crystal-field interactions associated with a half filled $4f^{7}$ shell.
The magnetic structure resolved here is of the $C$-AFM type, with partly frustrated first-neighbor ($J_{1}$) spin interactions. Based on this
result and previous studies on other members of the $R_{m}$(Co,Rh,Ir)$_{n}$In$_{3m+2n}$ family, we conclude that the relative
orientation between neighboring $R$-spins is determined primarily from a close competition between first- ($J_{1}$) and second-nearest-neighbor 
($J_{2}$) AFM exchange interactions. This may be an important factor behing the complex magnetic behavior
of the Ce-based compounds. The critical behavior close to the magnetic ordering transition in GdRhIn$_{5}$ was also investigated.
The magnetic order parameter shows a power-law behavior close to the N{\' e}el temperature,
characteristic of a second-order transition such as in Gd$_{2}$IrIn$_{8}$.\cite{Granado}
A theoretical renormalization group analysis was performed.
Our results suggest that spin anisotropy terms, possibly arising from dipolar and other fairly weak interactions,
must be claimed to understand the second order of the paramagnetic transition of GdRhIn$_{5}$ and the related compound
Gd$_{2}$IrIn$_{8}$. 

\section{Experimental Details}

A single crystal of GdRhIn$_{5}$ was grown by the In-flux method as described previously.\cite{Pagliuso,Fisk}
The studied surface was finely polished with Al$_{2}$O$_{3}$ powder, yielding a single-peaked
mosaic structure of $\sim 0.02 ^{\circ}$ full width at half maximum (FWHM). 
The x-ray diffraction measurements were performed on the XRD2
beamline, placed after a dipolar source at the Laborat\'{o}rio
Nacional de Luz S\'{i}ncrotron, Campinas, Brazil.\cite{Giles} The
sample was mounted on the cold finger of a commercial closed-cycle
He cryostat with a cylindrical Be window. The cryostat was fixed
onto the Eulerian cradle of a commercial 4+2 circle
diffractometer, appropriate for single crystal x-ray diffraction
studies. The energy of the incident photons was selected by a
double-bounce Si(111) monochromator, with water-refrigeration in
the first crystal, while the second crystal was bent for sagittal
focusing. The beam was vertically focused or collimated by a bent
Rh-coated mirror placed before the monochromator, which also
provided filtering of high-energy photons (third and higher order
harmonics). Unless otherwise noted, a vertically focused beam was
used in our measurements, delivering, at 7.24 keV, a flux of
$3\cdot 10^{10}$ photons/s at 100 mA in a spot of $\sim 0.6$ mm
(vertical) x $2.0$ mm (horizontal) at the sample, with an energy
resolution of $\sim 5$ eV. Our experiments were performed in the
vertical scattering plane, i.e., perpendicular to the linear
polarization of the incident photons. In most measurements, a
solid state detector was used, except in the polarization study,
where a scintillation detector was placed after a Ge(111/333)
analyzer crystal. At the energy corresponding to the Gd $L_{II}$
edge, the analyzer placed at the Ge(333) reflection selects
($\sigma\rightarrow\sigma'$) scattering from the sample (i.e.,
scattered photons with the same polarization as the incident
photons), while Ge(111) does not significantly discriminate the
photon polarization
(($\sigma\rightarrow\sigma')+(\sigma\rightarrow\pi'$) channel).

\section{Magnetically Ordered Phase}

\subsection{Polarization and Resonance Properties}

Above $\sim 39$ K, all the observed Bragg peaks were consistent
with tetragonal symmetry of GdRhIn$_{5}$ (space group $P4/mmm$), without any detectable magnetic contribution.
Below $T_{N}=39$ K, additional ($h,k,l$) Bragg
reflections ($h$ integer; $k,l$ half-integers) were observed. Such
reflections were dramatically enhanced at the Gd $L_{II}$ and
$L_{III}$ edges ($E=7.93$ and 7.24 keV, respectively) due to
resonance phenomena (see below). Polarization analysis at the Gd $L_{II}$
edge using a Ge(111/333) analyser crystal
demonstrated pure ($\sigma\rightarrow\pi'$) scattering at these fractional positions at the
reciprocal space, showing that such reflections are magnetic in origin, with dipolar resonances at this
edge.\cite{Hill,Lovesey} 

\begin{figure}
\vspace{-1.0cm}
\includegraphics[width=0.95 \textwidth]{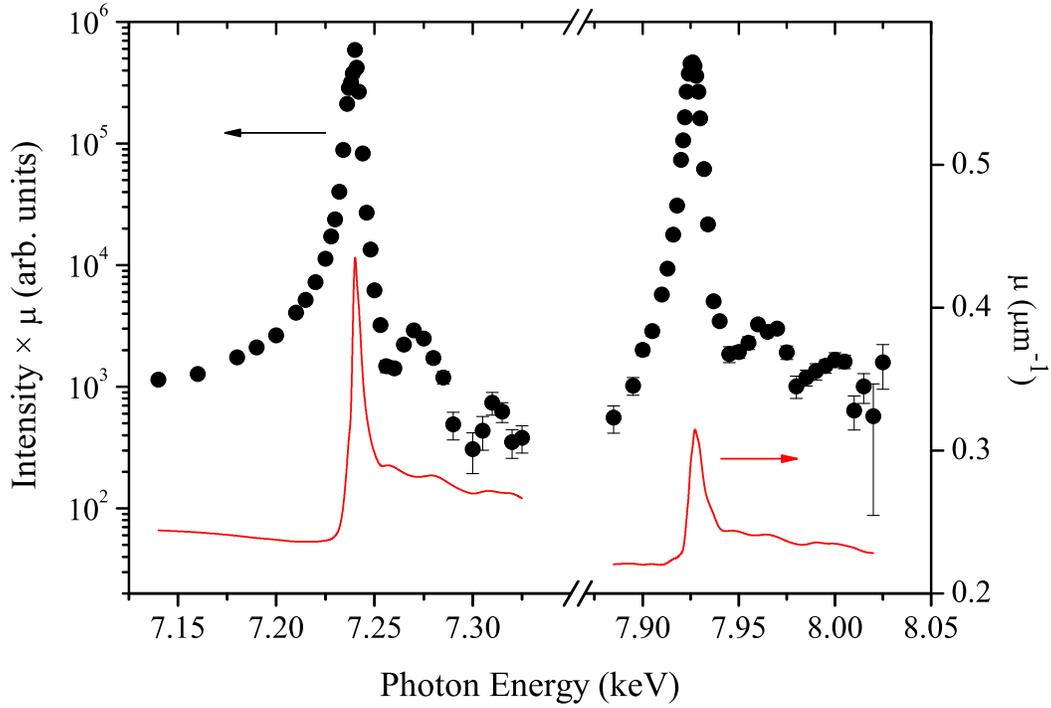}
\caption{\label{Energy} Energy-dependence of the integrated intensity of the (0,$\frac{1}{2},\frac{7}{2}$) Bragg
reflection (symbols) across the Gd $L_{III}$ and $L_{II}$ edges. Data were corrected
by the absorption coefficient $\mu$ obtained from the fluorescence yield (solid lines).}
\vspace{1.0cm}
\end{figure}

The energy-dependence of the absorption-corrected intensity of the (0,$\frac{1}{2},\frac{7}{2}$)
magnetic Bragg reflections at $\sim 12$ K are shown in Fig. \ref{Energy} around the
Gd $L_{III}$ and $L_{II}$ edges (filled symbols).
The energy-dependencies of the absorption coefficient, $\mu(E)$,
obtained from fluorescence emission, are also given as solid
lines. Resonant enhancements of over three orders of
magnitude were observed at both edges.
The intensity maximums
occur $\sim 2$ eV above the absorption edges, which were defined
as the inflection points of $\mu(E)$. This result is consistent
with a dominant dipolar nature ($2p\rightarrow5d$) for both
resonances. Intensity oscillations of the (0,$\frac{1}{2},\frac{7}{2}$)
magnetic peak were also observed above the edges, which we ascribe
to a magnetic diffraction anomalous fine structure
(DAFS).\cite{Stragier}

\subsection{Magnetic structure}

\begin{table}
\caption{\label{Direction} Comparison between observed and calculated
intensities of magnetic Bragg reflections at 12 K, normalized by the most intense
reflection, assuming the moments $\vec{m}$ along each one of
the three axis of the unit cell. Experimental data were taken on resonance conditions, 2 eV above
the Gd $L_{II}$ edge.}
\begin{ruledtabular}
\begin{tabular}{llllll}
($h,k,l$) & $I_{obs}$ & $\vec{m} \parallel \vec{a}$ & $\vec{m}
\parallel \vec{b}$ & $\vec{m} \parallel \vec{c}$ \\
($0,-\frac{1}{2},\frac{7}{2}$) & 96(3) & 100 & 11 & 21 \\
($0,\frac{1}{2},\frac{7}{2}$) & 100 & 100 & 11 & 21 \\
($0,-\frac{3}{2},\frac{7}{2}$) & 68(2) & 93 & 100 & 21 \\
($0,\frac{3}{2},\frac{7}{2}$) & 68(2) & 93 & 100 & 21 \\
($0,-\frac{1}{2},\frac{9}{2}$) & 99(2) & 90 & 11 & 35 \\
($0,\frac{1}{2},\frac{9}{2}$) & 96(2) & 90 & 11 & 35 \\
($0,-\frac{3}{2},\frac{9}{2}$) & 83(2) & 83 & 100 & 35 \\
($0,\frac{3}{2},\frac{9}{2}$) & 79(2) & 83 & 100 & 35 \\
($0,-\frac{1}{2},\frac{11}{2}$) & 95(2) & 77 & 11 & 53 \\
($0,\frac{1}{2},\frac{11}{2}$) & 90(2) & 77 & 11 & 53 \\
($1,-\frac{1}{2},\frac{11}{2}$) & 40(1) & 41 & 21 & 100 \\ 
($1,\frac{1}{2},\frac{11}{2}$) & 43(1) & 41 & 21 & 100 \\

\end{tabular}
\end{ruledtabular}
\vspace{1.0cm}
\end{table}

\begin{figure}
\includegraphics[width=0.95 \textwidth]{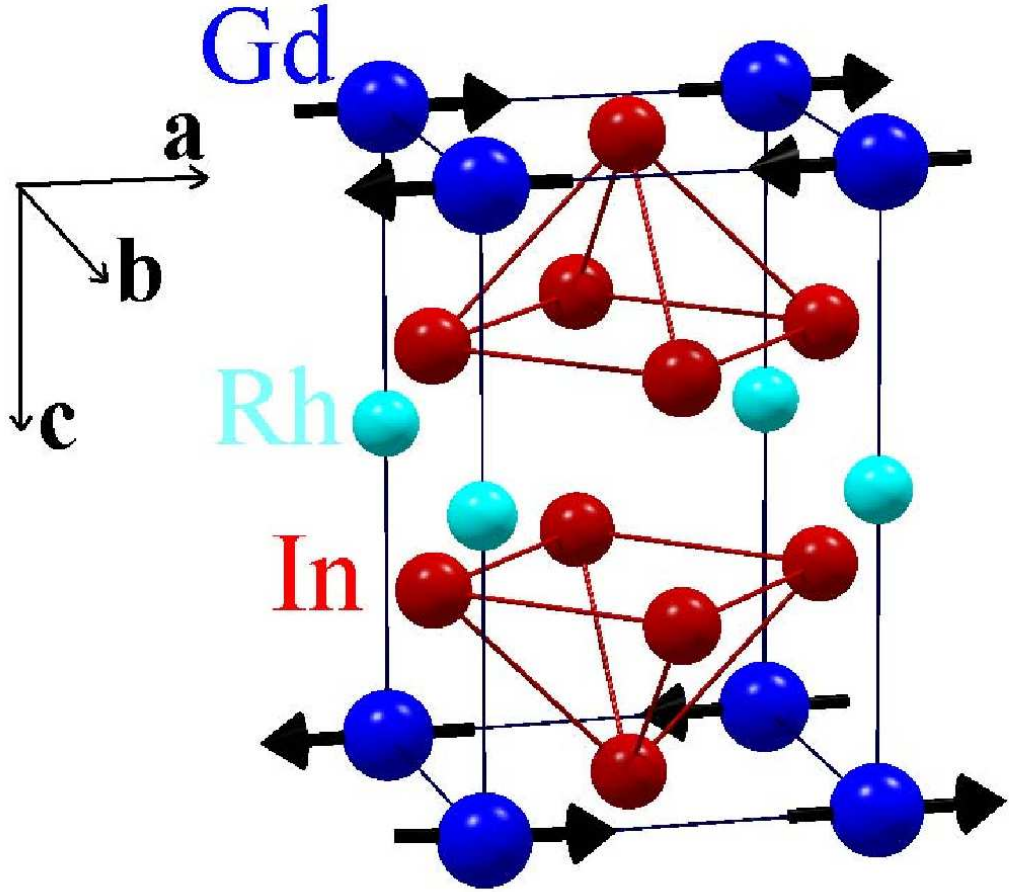}
\vspace{-1.0cm}
\caption{\label{Magneticstructure} Magnetic structure of GdRhIn$_{5}$.}
\vspace{1.0cm}
\end{figure}

The positions in reciprocal space where magnetic Bragg reflections were
observed lead to AFM structures with propagation vector
$\vec{\tau}=(0,\frac{1}{2},\frac{1}{2}$). Since there is a single magnetic ion per chemical unit cell
in the tetragonal structure of GdRhIn$_{5}$,
the relative neighboring spin orientations are unequivocally determined
from the $\vec{\tau}$-vector. According to this, linear ferromagnetic chains 
along the $\vec{a}$-direction are antiferromagnetically-coupled along the other two axes (i.e., a $C$-type AFM structure).
The direction of the magnetic moments may be obtained from the 
intensities of some magnetic Bragg peaks.
The expression for such intensities shows a simple form in the present case for dipolar
resonances and colinear magnetic structures, and is given by
$I^{M}(\vec{\tau})\propto(\vec{m}\cdot \vec{k_{s}})^2$, where $\vec{m}$ is the magnetic moment,
and $\vec{k_{s}}$ is the wave vector of the scattered light.\cite{Hill,Lovesey}
Comparisons of the observed intensities with calculated ones for the magnetic moments along
$\vec{a}$, $\vec{b}$, and $\vec{c}$-directions are given in Table \ref{Direction} for both compounds.
Good agreement between calculated and experimental data are obtained for the magnetic
moments pointing towards the direction of the ferromagnetic chains, i.e., the $\vec{a}$-direction.
Figure \ref{Magneticstructure} displays the magnetic structure of GdRhIn$_{5}$, as determined above.
This structure is consistent with a $^{155}$Gd M\"{o}sbauer spectroscopy study in this compound, which indicated
a colinear magnetic structure with the spins lying in the {\it ab} plane.\cite{Latka}

\subsection{Search for symmetry lowering of the crystal or electronic structure}

It is interesting to note that the magnetic structure shown in Fig. \ref{Magneticstructure} does not have a
tetragonal symmetry, in the sense that the ferromagnetic chains are aligned along one specific
direction in the {\it ab} plane, namely the $\vec{a}$-direction. It is therefore valid to ask whether a symmetry lowering
of the crystal and/or electronic structure from the $P4/mmm$ space group also occurs, either by an orthorhombic distortion
of the lattice parameters or by the presence of a charge density wave. These possibilities were also investigated by synchrotron
x-ray diffraction.

Firstly, a hypothetical orthorhombic distortion was probed by measuring the Bragg interplane distances $d_{hkl}$ of a set of 
$(00l)$, $(h0l)$, and $(0kl)$ charge reflections,
using a monochromatic x-ray beam with energy $E=14472$ eV  and resolution $(\delta E / E = 1.0 \times 10^{-3})$. The irradiated (001) surface was
the same used in the magnetic diffraction investigation. A 
scintillation detector was placed after a Si(111) analyzer crystal. We used the usual relation $d_{hkl}^2=1/[(h/a)^{2}+(k/b)^{2}+(l/c)^{2}]$ for orthogonal
axes, obtaining
the lattice parameter $c$ directly from $d_{00l}$. The $a$ and $b$ parameters were then obtained from $d_{h0l}$ and $d_{0kl}$.
Table \ref{lattice} shows the results at 20 K and 300 K, indicating a metrically tetragonal phase within experimental errors.

\begingroup
\squeezetable
\begin{table*}
\caption{\label{lattice} Extraction of the $a$, $b$, and $c$ lattice parameters at 20 K and 300 K from
selected $(00l)$, $(h0l)$, and $(0kl)$ Bragg peaks. The width of radial ($\theta-2\theta$) scans are also given
in terms of $\delta \theta/tan(\theta)$ (full width at half maximum), were $\theta$ is the Bragg angle.}
\begin{ruledtabular}
\begin{tabular}{c c c c c} 
Miller indices & $a$, $b$, or $c$ & $\delta \theta/tan(\theta)$ (degrees) & $a$, $b$, or $c$ & $\delta \theta/tan(\theta)$ (degrees) \\
 & 20 K & 20 K & 300 K & 300 K \\
\hline \\
(006) & $c=7.4302(3)$ \AA & 0.0364(5) & $c=7.4479(4)$ \AA & 0.0367(6) \\
(004) & $c=7.4299(3)$ \AA & 0.0372(6) & $c=7.4473(4)$ \AA & 0.0377(6) \\
(308) & $a=4.5919(8)$ \AA & 0.0370(8) & $a=4.6066(8)$ \AA & 0.0378(9) \\
(207) & $a=4.5922(8)$ \AA & 0.0363(8) & $a=4.6059(8)$ \AA & 0.0369(8) \\
(044) & $b=4.5935(6)$ \AA & 0.0373(8) & $b=4.6078(6)$ \AA & 0.0386(8) \\
(043) & $b=4.5934(6)$ \AA & 0.0359(8) & $b=4.6081(6)$ \AA & 0.0376(8) \\
\end{tabular}
\end{ruledtabular}
\end{table*}
\endgroup

The possibility of an orthorhombic structure presenting a mosaic of domains with interchanged $\vec{a}$ and $\vec{b}$ axes was also considered. In this case,
a two-peak structure is expected at each $(h0l)$ or $(0kl)$ reflection.
Such a structure was not observed in our measurements, rather
yielding single symmetric
Lorentzian line shapes. In this case, the upper limit for the orthorhombic distortion is set by the measured peak widths. This information is also given
in Table \ref{lattice} in terms of $\delta \theta / tan(\theta)$. A larger $\delta \theta / tan(\theta)$ for $(h0l)$ and $(0kl)$
reflections with respect to $(00l)$ reflections would be consistent with an orthorhombic distortion or anisotropic strain in the $ab$ plane.
Nonetheless, it is seen that no additional broadening of the $(h0l)$ or $(0kl)$ peaks with respect to $(00l)$ was observed
within our resolution.
In addition, the peak widths are the same at 20 K and 300 K, showing the absence of any temperature-dependent distortion or strain.
Using the data of Table \ref{lattice}, the upper limit of the hypothetical orthorhombic distortion is inferred to be
($(b-a)/a) \lesssim 2 \times 10^{-4}$ for both single-domain or mosaic (multi-domain)
distortions. We conclude that no relevant bulk lattice distortion or anisotropic strain
associated with the anisotropy of the magnetic structure along the $\vec{a}$-axis take place in this compound.

The possible presence of charge density waves (CDWs) was probed by
a systematic search in reciprocal space at x-ray energies of 7930 eV (resonant condition at Gd $L_{II}$ absorption edge) and
at 7106.7 eV (below the Gd $L_{III}$ absorption edge).
A set of charge reflections - (001), (002), (011), (022), (021), (102) - was chosen to define the borders
of the one-dimensional scans in reciprocal space. 
Several scans along high symmetry $h$, $k$, and $l$ mixed and unmixed directions were performed at $T=11.5$ K.
The search was completed by two-dimensional $h-k$, $h-l$ and $k-l$ maps with a common border at the (022) reciprocal space position
($\delta h,k,l = 0.5$). We did not
find any extra peak that might be assigned to previously unknown ordered structures. Although our scan procedure had been unable to directly reveal the
existence of new electronic or structural phases in GdRhIn$_{5}$, we cannot completely rule out the possibility of existence of low-symmetry incommensurate phases such as CDWs,
since we did not investigate the whole reciprocal space volume.

\subsection{Discussion: Rise of the $C$-AFM magnetic structure and competition with $G$-AFM: the role of long-range
exchange interactions}

\begin{figure}
\includegraphics[width=0.95 \textwidth]{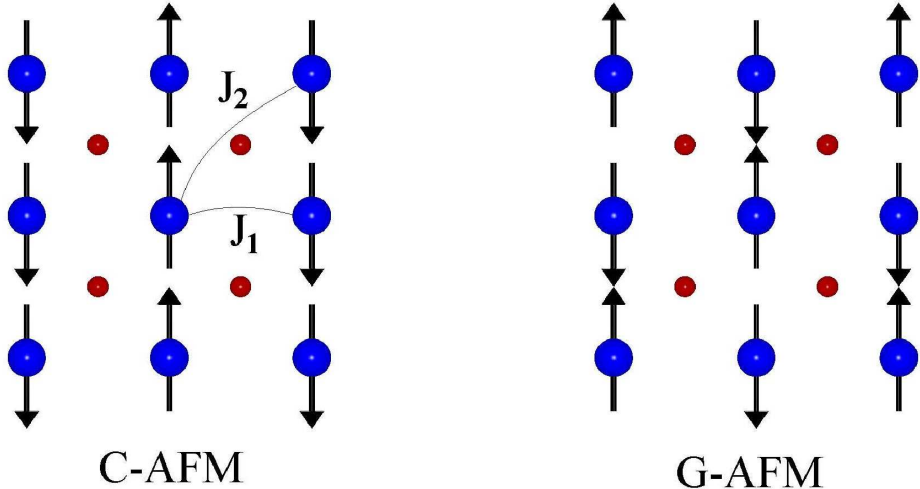}
\vspace{-1.0cm}
\caption{\label{drawing} (Color online) Two-dimensional representation of the $C$-AFM and $G$-AFM spin structures.
The first- and second-nearest-neighbor exchange interactions ($J_{1}$ and $J_{2}$, respectively)
are also indicated.}
\vspace{1.0cm}
\end{figure}

In the $R_{m}$(Co,Rh,Ir)$_{n}$In$_{3m+2n}$ ($R=$ rare earth $\neq$
Ce) family, the main coupling between the $R$ ions comes from the 
Ruderman-Kittel-Kasuya-Yosida (RKKY) mechanism.\cite{RKKY} If we
consider antiferromagnetic couplings only between the first-($J_{1}$)
and the second-nearest-neighbors ($J_{2}$), the magnetic ground state
is determined by their relative strength $\alpha\equiv J_{1}/J_{2}$. 
The $G$-AFM structure represented in Figure \ref{drawing} frustrates
$J_{2}$ but satisfies $J_{1}$ and would be favored by $\alpha>>1$.
On the other hand, the $C$-AFM structure satisfies part of $J_{1}$
and all $J_{2}$ interactions and thus may be favored if $J_{1}$ and $J_{2}$ are comparable in magnitude. 
For the RKKY interaction, $\alpha$ usually depends on the first-
and second-neighbor distances among the rare-earth ions and on the
topology of the Fermi surface of each compound. However, in these
compounds the topology and volume of the Fermi surface depends 
little on the particular rare-earth ion since, except for the Ce based
compounds, the localized $4f$ electrons do not hybridize considerably
with the conduction band. Also, we note that the tetragonal $a$-lattice 
parameter shows only small variations along the $R_{m}$(Co,Rh,Ir)$_{n}$In$_{3m+2n}$
series ($\lesssim 2$ \%).\cite{Pagliuso} It is therefore not completely
surprising that the magnetic structures of all known $R_{m}$(Co,Rh,Ir)$_{n}$In$_{3m+2n}$ 
($R=$ Nd to Gd) compounds are similar, characterized
by ferromagnetic chains along a specific first-nearest-neighbor direction
($\vec{a}$) with an AFM coupling along $\vec{b}$ and $\vec{c}$,
i.e., the $C$-AFM structure (see Fig. \ref{Magneticstructure} and
refs. \cite{Mitsuda,Amara,Chang,Granado}).

The orientation of the magnetic moment in any of these magnetic structures
is determined by anisotropic interactions. The direction of the staggered
moment on GdRhIn$_{5}$ is in agreement with other tetragonal compounds
with the same antiferromagnetic wave-vector and staggered moment direction,
such as GdAu$_{2}$Si$_{2}$ \cite{rotter144418} and GdCu$_{2}$Si$_{2}$.\cite{Barandiarana88} 
It has been argued that in theses cases the magnetic
dipolar interaction is the dominant source of anisotropy (being of
the order of tens of $\mu eV$).\cite{rotter144418} The same appear to hold for GdRhIn$_{5}$.
Nonetheless, we should mention that other possible sources of anisotropy for Gd compounds,
such as a spin-orbit coupling of the conduction electrons\cite{Tosti} or crystal electric field
via excited states\cite{Wybourne,Barnes1,Barnes2,Urbano} might in principle be relevant to the problem.

The same general conclusions for the ground state may remain valid when 
the spin anisotropy due to the relevant crystal field effects is introduced 
in the Hamiltonian for $R \neq$ Gd. In this case, the crystal field effects determine the spin direction with 
respect to the unit cell axes and may affect $T_{N}$,\cite{Pagliuso5} but the relative 
orientation between neighboring $R$-spins is still determined mainly by $\alpha$.

The extension of the above scenario for the Ce-based compounds is
not straightforward. This is because the Ce $4f$ electrons may be
hybridized with the conduction band. In cases where the Ce $4f$ electrons
are itinerant, the RKKY mechanism is no longer applicable. On
the other hand, for compounds with localized $4f$ moments, the above
scenario of $J_{1}/J_{2}$ competition might be useful. Particularly,
deHaas van Alphen measurements on Ce$_{1-x}$La$_{x}$RhIn$_{5}$ 
showed no significant change in the Fermi surface topology or volume
over the entire doping range ($0<x<1$),\cite{Alver} showing that the $4f$
electrons remain localized. Thus, we focus our discussion on the magnetic
structure of CeRhIn$_{5}$ with localized moments. This 
compound shows AFM alignment along two directions ($\vec{a}$ and
$\vec{b}$), and a spiral alignment along the tetragonal $\vec{c}$-axis,\cite{Bao1}
defining a propagation vector $\vec{\tau}=(\frac{1}{2},\frac{1}{2},0.297)$. This magnetic 
structure may be seen as an intermediate case between the $G$-AFM
structure ($\vec{\tau_{1}}=(\frac{1}{2},\frac{1}{2},\frac{1}{2}))$ and a $C$-AFM structure
with the FM chains along the $\vec{c}$-direction ($\vec{\tau_{2}} =(\frac{1}{2},\frac{1}{2},0)$).
Notice that the $C$-AFM structure is expected to be a competitive ground state, 
since $\alpha$ is expected to be similar for CeRhIn$_{5}$ and GdRhIn$_{5}$ due to presumably similar Fermi surfaces.
On the other hand, the $G$-AFM state also appears to be competitive
for the Ce-based compounds, since it is the ground state of CeIn$_{3}$\cite{Benoit,Lawrence} 
and Ce$_{2}$RhIn$_{8}$.\cite{Bao2} It is therefore not implausive
to infer that the incommensurate magnetic structure of CeRhIn$_{5}$
is actually a result of a close competition between the $G$-AFM and
$C$-AFM ground states, paving the way for the stabilization of an 
intermediate spiral phase, perhaps with the aid of very long-range
RKKY interactions ($J_{3}$, $J_{4}$, etc).

\begin{figure}
\includegraphics[width=0.5 \textwidth]{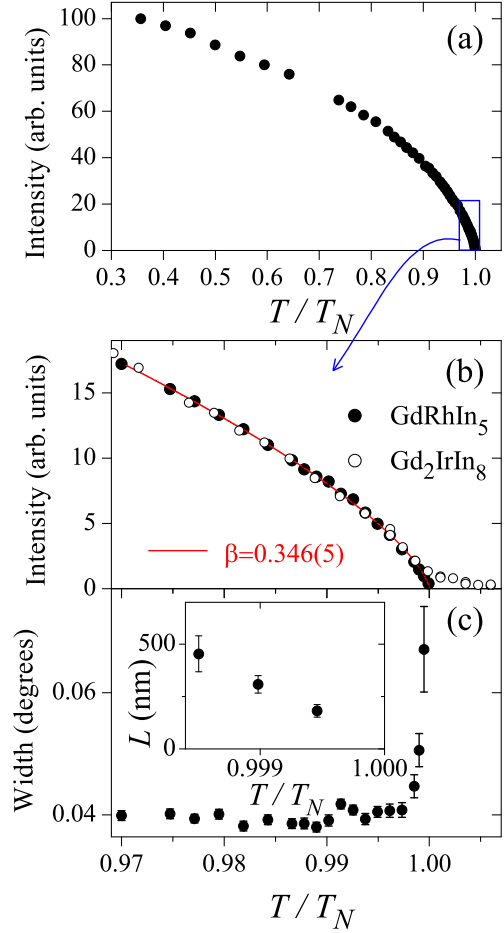}
\vspace{-1.0cm}
\caption{\label{CriticalGd115} (Color online) Temperature-dependence of the (a,b) integrated intensity and (c) width of radial
($\theta - 2\theta$) scans of the
($0,\frac{1}{2},\frac{11}{2}$) magnetic Bragg reflection of GdRhIn$_{5}$ and ($\frac{1}{2},0,4$) reflection of
Gd$_{2}$IrIn$_{8}$ (taken from ref. \cite{Granado}). A fit to a critical power law behavior, $(T_{N}-T)^{2\beta}$, characteristic of a second-order
transition, is given in (b) as a line. The inset show the correlation length $L$, obtained
from the data in (c) after deconvolution of the instrumental width. The experiment was performed in resonance
conditions, 2 eV above the Gd $L_{II}$-edge.}
\vspace{1.0cm}
\end{figure}

\section{Critical Behavior}

\subsection{Experimental}

The critical behavior of the sublattice magnetization at the paramagnetic transition was investigated.
Figures \ref{CriticalGd115}(a,b) show the $T$-dependence of the intensity of the ($0,\frac{1}{2},\frac{11}{2}$)
magnetic Bragg peak of GdRhIn$_{5}$ in different $T$-intervals. Close to and below $T_{N}=39$ K, this could be fitted
by a power-law behavior, $I \propto (1-T/T_{N})^{2\beta}$, which is characteristic of a second-order transition.
The experimental determination of the critical parameter $\beta$ depends slightly on the
temperature interval in which the fitting is performed. For fits in $T$-intervals $(T_{N}-T)/T_{N}
< 0.01$, $< 0.03$, and $< 0.05$, one obtains $\beta= 0.370(16), 0.346(5)$, and $0.339(4)$, respectively. Quoted errors in
parentheses are statistical only, and represent one standard deviation. Figure \ref{CriticalGd115}(c) shows the $T$-dependence
of the width of the ($0,\frac{1}{2},\frac{11}{2}$) magnetic peak, as obtained in radial ($\theta-2\theta$) scans.
Much below $T_{N}$, the width is instrumental only, indicating long-range order with correlation length above $\sim 1000$ \AA.
For $0.998T_{N}<T<T_{N}$, a peak broadening above the instrumental resolution was noticed. The inset of
Fig. \ref{CriticalGd115}(c) displays the magnetic correlation length in this $T$-interval, obtained from the
peak broadening data. We should mention that, above $T_{N}$, the intensities were below our
detection limit for this sample, thus the short-range dynamic correlations in the paramagnetic phase could not be investigated.

These results for GdRhIn$_{5}$ may be compared to the previously reported measurements
for Gd$_{2}$IrIn$_{8}$, where a larger critical parameter $\beta=0.39$ was obtained for the critical interval
$(T_{N}-T)/T_{N} < 0.10$.\cite{Granado} The comparison becomes clear in Fig. \ref{CriticalGd115}(b), where the critical
behavior of Gd$_{2}$IrIn$_{8}$, using data of ref. \cite{Granado}, is directly compared to GdRhIn$_{5}$. In the ordered
phase ($T < T_{N}$), both compounds appear
to show identical behavior, described by the same critical exponent $\beta \sim 0.35$. However, while the magnetic 
intensities tends to zero as $T \sim T_{N}$ for GdRhIn$_{5}$, a significant residual scattering was observed near and
above the transition for Gd$_{2}$IrIn$_{8}$, which may be mostly ascribed to magnetic correlations in the near surface
region due to long-range correlated quenched disorder.\cite{Granado,Altarelli} This effect smooths out the transition observed by
x-ray diffraction, and interfere severely
in the extraction of the $\beta$ critical exponent. We conclude that, although both compounds were equally finely polished before measurements,
our studied Gd$_{2}$IrIn$_{8}$ surface showed a larger degree of near-surface disorder, leading
to a less reliable extraction of the $\beta$ exponent. Perhaps the most relevant information from this analysis is that
a direct comparison of the magnetic intensities of Gd$_{2}$IrIn$_{8}$ and Gd$_{2}$RhIn$_{5}$ near $T_{N}$ reveals an identical
behavior in the $T$-interval where the surface disorder effects are negligible, consistent with an identical critical exponent
$\beta \sim 0.35$. This conclusion is consistent with the prediction that both compounds belong to the same universality
class for magnetism (see below).

\subsection{Renormalization Group Analysis}

The experimental critical behavior described above may
be compared with theoretical expectations based on the symmetry of
the crystal and magnetic structures of this compound. As demonstrated
by Mukamel \emph{et al.},\cite{Mukamel1976,Mukamelb1976} the Ginzburg-Landau-Wilson
(GLW) Hamiltonian for the class of antiferromagnetic problems is written
in terms of a staggered order parameter (OP) with a total number of
components $N=nm$, where $n$ is the number of spin components allowed
by the irreducible representations of the paramagnetic space group,
or equivalently, the number of degrees of freedom for the critical
spin fluctuations, and $m$ is the number of ways the unitary cell
may be enlarged by all distinct AFM ordering wave-vectors allowed by
the symmetry of the crystal. The total number of components may easily
exceed $N\geq4$,\cite{Brezin1974} opening up the possibility to
classical fluctuation-induced first-order transitions, such as those already
reported experimentally.\cite{Mukamel1976,Mukamelb1976} The criterion
for the study of phase transitions using renormalization group (RG)
theoretical methods is based on the stability of the fixed points
in the RG flow, in the sense that the phase transition is of second
order when the flow continuously approaches a fixed point which is
stable with respect to the fluctuations of the staggered field, and
indicates something else like an abrupt or an smeared transition otherwise,
when the flow exhibits a runaway.\cite{Bak1976}

Because of the critical exponents universality, the phase transition
is sensitive only to a few parameters like the dimension $d$ of the
system, the number of components $N$ of the OP and the symmetry of
the crystal, which reflects in the anisotropies of the Hamiltonian.
As shown by Br\'{e}zin \emph{et al.},\cite{Brezin1974} for $N\geq4$,
the isotropic fixed point is always unstable with respect to anisotropies
in the original Hamiltonian, while for $N<4$ this fixed point is
always stable because the O$(N)$ symmetry is dynamically generated
near the critical point.


Here, we concentrate the analysis on a few compounds of the Gd$_{m}$(Rh,Ir)$_{n}$In$_{3m+2n}$ 
series, including the title compound. In the cubic GdIn$_{3}$ crystal,
the observed AFM propagation vector $\vec{\tau}=(\frac{1}{2},\frac{1}{2},0)$
\cite{Malachias} represents an ordering state which is degenerated
with the states represented by the propagation vectors $(0,\frac{1}{2},\frac{1}{2})$
and $(\frac{1}{2},0,\frac{1}{2})$, resulting in $m=3$ distinct AFM
propagation vectors allowed by the crystal symmetry. Since $n\leq3$ is the
total number of spin degrees of freedom, the staggered OP of GdIn$_{3}$
has a total number of $N=3n\leq9$ components. In the tetragonal Gd$_{2}$IrIn$_{8}$
crystal, the cubic symmetry is broken in the {[}001{]} direction by
non-magnetic planes of Ir.\cite{Granado} This crystal orders antiferromagnetically
along the $\vec{\tau}=(\frac{1}{2},0,0)$ direction, which is equivalent
by symmetry to the $(0,\frac{1}{2},0)$ direction, giving $m=2$ and
$N=2n\leq6$ OP components. Since the magnetization of GdRhIn$_{5}$
orders along the $\vec{\tau}=(0,\frac{1}{2},\frac{1}{2})$ direction
(which is equivalent by symmetry to the $(\frac{1}{2},0,\frac{1}{2})$
direction), its staggered OP has the same number of components of
GdIr$_{2}$In$_{8}$, $N=2n\leq6$.

In isotropic magnetic systems, the equality $n=3$ holds for the number of spin degrees of freedom,
and all above relations for $N$ are valid with the equality sign.
We conclude that if GdRhIn$_{5}$, Gd$_{2}$IrIn$_{8}$, and GdIn$_{3}$ were perfectly isotropic spin systems,
first-order transitions would be obtained for all three compounds,\cite{Brezin1974} in contrast with the second-order
transitions observed for GdRhIn$_{5}$ (see Fig. \ref{CriticalGd115}) and Gd$_{2}$IrIn$_{8}$ (see ref. \cite{Granado}).
We conclude that spin anisotropy must be properly taken into account for a correct analysis of the critical behavior
or these compounds.
As noted in Section III.D, a number of distinct possible sources of anisotropy may be anticipated for
Gd compounds with half-filled $4f^{7}$ shell, such as dipolar interactions,\cite{rotter144418} spin-orbit coupling of the
conduction electrons\cite{Tosti} or crystal electric field via excited states.\cite{Wybourne,Barnes1,Barnes2,Urbano}
The strength of such interactions is typically of the order of tens of $\mu$eV.\cite{rotter144418,Tosti} Even though
this energy scale is about three orders of magnitude weaker than typical exchange energies in Gd systems, we
note, for example, that there are consistent evidences that in Gd metal the dipolar
interaction is not only responsible for the ground state anisotropy but also determines
its critical behavior.\cite{Hennenberger99,Srinath99}

In order to proceed with our analysis, the major source of anisotropy must be identified.
Since there are recent indications
that dipolar interactions are the major source of anisotropy \cite{rotter144418}
and responsible for the specific heat behavior in an extensive variety
of Gd compounds,\cite{Rotter2001,Bouvier1991,Blanco1991} we 
pay attention to the possible influence of the dipolar coupling in
the critical behavior of these materials. The dipolar anisotropy breaks
the rotational symmetry of the Gd spins by lowering the size of the
space of degenerated states where the spins are allowed to fluctuate.
In the RG sense, the influence of the dipolar interaction will be
decisive if it proves to be a relevant source of anisotropy in a previously
isotropic Hamiltonian.\cite{aharoniI} The difficulty of the RG method
here is that it leads to rather inconclusive results when the RG flow
has no stable fixed points, since the flow rapidly moves towards a
region where the technique is no longer valid. To see this, we write
down the most general classical Hamiltonian that describes the physics
of the isotropic Gd spin problem, which corresponds to a GLW Hamiltonian
of $m$ coupled O$(n)$ symmetric models,\begin{eqnarray}
H_{0}(\phi) & = & \int\textrm{d}^{d}x\left\{ \frac{1}{2}\sum_{\alpha,i}\left[r_{0}\phi_{\alpha i}^{2}+\left(\nabla\phi_{\alpha i}\right)^{2}\right]\right.\nonumber \\
 &  & \qquad\quad+u\sum_{\alpha,i,j}\phi_{\alpha i}^{2}\phi_{\alpha j}^{2}+v\sum_{\alpha\neq\beta}\sum_{ij}\phi_{\alpha i}^{2}\phi_{\beta j}^{2}\nonumber \\
 &  & \qquad\quad\left.+w\sum_{\alpha\neq\beta}\sum_{ij}\phi_{\alpha i}\phi_{\beta i}\phi_{\alpha j}\phi_{\beta j}\right\} ,\label{O(n)H}\end{eqnarray}
 where $\alpha,\beta=1,...,m$ indexes the distinct AF wave-vectors
and $i,j=1,...n\leq3$ labels the spin components in a given orthogonal
basis, like $x,y,z$. Note that all terms are written as powers of
scalar products of the spin components $\vec{\phi}_{\mu}\cdot\vec{\phi}_{\nu}$
because of the assumed rotational symmetry of the spins near the phase
transition. The quartic terms differ only by the different ways to
combine the Greek indexes that label the equivalent AF wave-vectors.
Including a general dipolar interaction term, which in the antiferromagnetic
case has the form,\cite{aharoniI}\begin{equation}
H_{D}(\phi )=\sum_{\alpha i}\left[(\nabla\phi_{i\alpha})^{2}-f(\partial_{i}\phi_{i\alpha})^{2}+h\sum_{j}\partial_{i}\phi_{i\alpha}\partial_{j}\phi_{j\alpha}\right],\label{DipolarH}\end{equation}
 where $f$ and $h$ are proportional to the dipolar coupling constant
$\left(g\mu_{B}\right)^{2}$, with $g\mu_{B}$ being the total magnetic
moment of the Gd ion, we show in Appendix A 
that the total Hamiltonian\begin{eqnarray}
H & = & H_{0}+H_{D}\label{H}\end{eqnarray}
 has the isotropic fixed point for $N=nm<4$, and no stable fixed
points for $N>4$. In particular, the scaling of the dipolar parameters
from equation\begin{eqnarray}
\frac{\textrm{d}f}{\textrm{d}l} & = & f\textrm{e}^{-\eta_{f}},\nonumber \\
\frac{\textrm{d}h}{\textrm{d}l} & = & h\textrm{e}^{-\eta_{h}},\label{fl}\end{eqnarray}
 with\begin{eqnarray*}
\eta_{f}=\eta_{h} & = & \frac{32}{3}\left(\frac{1}{8\pi^{2}}\right)^{2}\left(2u^{2}+\left(m-1\right)\left(v^{2}+w^{2}\right)\right)\geq0\,,\end{eqnarray*}
 indicates that the dipolar interaction is irrelevant in the vicinity
of the isotropic fixed point for $N<4$, what corresponds in our case
($n=3$) to $m=1$. For larger $m$, however ($m=2$ for GdRhIn$_{5}$,
Gd$_{2}$IrIn$_{8}$ and $m=3$ for GdIn$_{3}$), this Hamiltonian
has no stable fixed points and, although this would point to the direction
of a fluctuation induced first order phase transition, it is not clear
what happens in this case. The fact that the $f$ and $h$ parameters
flow initially to zero is expected, since we do not include sources
of anisotropy in the bare Hamiltonian as the cubic term $(v)$ from
reference \cite{aharoniI}. Nevertheless, as pointed out in this reference,
such terms could well be self generated in a higher loop expansion.

From mean field calculations of the propagation vector and moment
direction in several Gd compounds, Rotter \emph{et al.}\cite{rotter144418}
have shown strong evidence that in these compounds the observed anisotropy
stems from the magnetic dipolar interaction. If we assume that dipolar
interaction is relevant at the phase transition, it must break the
spin rotational invariance of Hamiltonian (\ref{O(n)H}). Once this
symmetry is broken, the quartic terms should be written in the most
general way allowed by the crystallographic group. In the particular
case of the Gd series compounds we have studied, we have not identified
bi-critical, tri-critical or multi-critical points associated to the
paramagnetic phase transition, meaning that the spin fluctuations
are confined inside subspaces of degenerated spin configurations.
In other words, each of these subspaces in the spin space correspond
to an irreducible representation of the AF order parameter. The number
and the size of all the irreducible representations allowed follows
directly from the crystallographic point group symmetry and from the
position of the ordering wave-vector $\vec{\tau}$ in the Brillouin
zone (BZ). In the case of GdRhIn$_{5}$, the $P4/mmm$ space group
associated with the special point $\vec{\tau}=(\frac{1}{2},0,\frac{1}{2})$,
at the border of the BZ, has three irreducible representations of
dimension $m$, with the spin pointing along the unit-cell axes of
the crystal (for details, see Appendix B). In this case, the number
of spin degrees of freedom for each representation is just $n=1$.
The most general Hamiltonian would be\begin{eqnarray}
H & = & \frac{1}{2}\sum_{\alpha}\left[r_{0}\phi_{\alpha}^{2}+c\left(\nabla\phi_{\alpha}\right)^{2}\right]+u\sum_{\alpha}\phi_{\alpha}^{2}\phi_{\alpha}^{2}\nonumber \\
 &  & \qquad\qquad+v\sum_{\alpha\neq\beta}\phi_{\alpha}^{2}\phi_{\beta}^{2},\label{1DirrepH}\end{eqnarray}
 with the dipolar interaction in this case simply renormalizing the
gradient term. For $m=2$, as in the case of GdRhIn$_{5}$, this Hamiltonian has one $O(m)$ symmetric
fixed point\cite{Brezin1974} with $\beta=0.36$ in two-loop expansion.
The ordering wave vector $\vec{\tau}=(\frac{1}{2},0,0)$ of Gd$_{2}$IrIn$_{8}$
(also from space group $P4/mmm$) gives the same three irreducible
representations of dimension $m$, and therefore GdRhIn$_{5}$ and
Gd$_{2}$IrIn$_{8}$ both lie in the same universality class. In GdIn$_{3}$,
the $Pm3m$ group acting on a BZ with the ordering wave-vector $\vec{\tau}=(\frac{1}{2},\frac{1}{2},0)$
admits one irreducible representation of size $N=m$, with the spin
pointing along the $\vec{c}$-axis, and another one of size $N=2m$ with
the spins confined in the $ab$ plane. The Hamiltonian of first representation
follows the general form of Eq. (\ref{1DirrepH}), showing one isotropic
fixed point for $m<4$. The second representation is larger ($N=6$)
and its Hamiltonian follows Eq. (\ref{H}), where no stable fixed
points were found, which would indicate the possibility of a fluctuation-induced
first-order transition for this case, even if the anisotropy is included.

\section{Conclusions}

In summary, our resonant x-ray diffraction experiments show a $C$-AFM magnetic structure
for GdRhIn$_{5}$ with the spins lying along the FM chain direction. This structure is
rationalized in terms of a competition between first- and second-neighbors exchange interactions.
This scenario was extended to other members of the $R_{m}$(Co,Rh,Ir)$_{n}$In$_{3m+2n}$ ($R=$ rare earth)
family, in particular the Ce-based heavy-fermion superconductors. The critical behavior
close to the paramagnetic transition was investigated, revealing a second-order transition
with critical exponent $\beta \sim 0.35$ for GdRhIn$_{5}$ and Gd$_{2}$IrIn$_{8}$. A renormalization
group analysis predicts that both compounds belong to the same universality class, in agreement with experiment.
However, a first-order transition is predicted in the absence of spin anisotropy terms in the Hamiltonian, in contrast
to our results, indicating that such interactions may be important 
to stabilize a critical point in this family.
Indeed, we show that if dipolar interactions or any other relevant source of spin anisotropy that allows the Gd spins
to couple with the lattice is included, then
the renormalization group analysis predicts a second order phase transition for GdRhIn$_{5}$ and GdIr$_{2}$In$_{8}$ with $\beta= 0.36$ in two-loop
$\epsilon$-expansion, in good agreement with our experiments. 

\appendix

\section{RG procedure}

Since the dipolar interaction is marginally relevant at the one loop
level, the RG calculation has to go to a second loop expansion. Following
the standard RG procedure along the lines of reference,\cite{aharoniI}
we integrate out of the partition function the fluctuation modes with
wave-vectors $b^{-1}<\left|\mathbf{k}\right|<1$ (which we denote
by $\int_{>}dk$ and $b\gg1$) and rescale the spins. This leads to
a renormalization of the dipolar part of the Hamiltonian (\ref{DipolarH})
according to\begin{eqnarray}
\Gamma_{2}\left(\mathbf{q}\right) & = & -16\left(2u^{2}+\left(m-1\right)\left(v^{2}+w^{2}\right)\right)\times\nonumber \\
 &  & \int_{>}d\mathbf{k}G_{ij}\left(\mathbf{k}\right)\sum_{lm}I_{lmm}\left(\mathbf{k}-\mathbf{q}\right)\nonumber \\
 &  & -32\left(2u^{2}+\left(m-1\right)vw\right)\times\nonumber \\
 &  & \int_{>}d\mathbf{k}\sum_{m}G_{im}\left(\mathbf{k}\right)\sum_{l}I_{lmj}\left(\mathbf{k}-\mathbf{q}\right)\label{gamma2}\end{eqnarray}
 where\[
I_{\gamma\delta\beta}\left(\mathbf{p}\right)=\int_{b^{-1}<\left|\mathbf{k}\right|<1}\textrm{d}^{d}\mathbf{k}G_{\gamma\delta}\left(\mathbf{k}\right)G_{\gamma\beta}\left(\mathbf{k}+\mathbf{p}\right)\,,\]
 We assume that the dipolar interaction is strong enough in the critical
region, i.e $T-T_{c}\ll\frac{\left(g\mu_{B}\right)^{2}}{a^{3}}$,
so that the bare propagator is given by\[
G_{ij}\left(\mathbf{k}\right)=\frac{1}{k^{2}}\left[\delta_{ij}-h_{0}\frac{k_{i}k_{j}}{k^{2}}+f_{0}\left(\frac{k_{i}}{k}\right)^{2}\delta_{ij}\right]\,,\]
 with $f_{0}$ and $h_{0}$ being proportional to the dipolar coupling
constant and to geometric factors reflecting the lattice symmetry
(see eq. 23 from ref.\cite{aharoniI}). After performing the integrals
on equation (\ref{gamma2}) and expanding $\Gamma_{2}\left(\mathbf{q}\right)$
to second order in $\mathbf{q}$, which though rather cumbersome is
straightforward, we obtain equation (\ref{fl}). Since $f$ and $h$ flow
initially to zero, the available fixed points up to this order correspond
to the fixed points of the isotropic Hamiltonian (\ref{O(n)H}). The
RG of the isotropic problem has been studied in detail by Ref. \cite{Aharony1975},
which indicates the absence of stable fixed points for $N=3m>4$.
The absence of stable fixed points within the isotropic model remains inaltered with the inclusion of long-range correlated quenched
disorder.\cite{Halperin}
In one-loop at least, the disordered isotropic model produces no new
stable fixed points \cite{Uchoa} beyond the two unphysical fixed
points previously found by Halpering and Weinrib.\cite{Halperin}

\section{Irreducible representations of $\textrm{GdRhIn}_{5}$}

The simplest way to obtain all the irreducible representations allowed
by the crystallographic point group for a given ordering point $\vec{\tau}$
in the BZ is to decompose the spins in a given basis \emph{fixed} with
respect to $\vec{\tau}$, and then apply all the point group symmetry
operations to see how the spin components in the original basis will
change. The idea is that if we properly choose the original spin basis,
then we are able to identify the subspaces where the spins will be
confined by the application of the symmetry operations allowed by
the crystal point group only. For GdRhIn$_{5}$, which ordering wave-vectors
are\begin{eqnarray*}
\vec{\tau}_{\alpha=1} & = & (\frac{1}{2},0,\frac{1}{2})\\
\vec{\tau}_{\alpha=2} & = & (0,\frac{1}{2},\frac{1}{2})\end{eqnarray*}
 we will define the spin basis by $\{\vec{\nu}_{1,\alpha},\vec{\nu}_{2,\alpha}\vec{\nu}_{3,\alpha}\}$,
where\begin{eqnarray*}
\vec{\nu}_{1,\alpha=1} & = & (1,0,0)\\
\vec{\nu}_{2,\alpha=1} & = & (0,1,0)\\
\vec{\nu}_{3,\alpha=1} & = & (0,0,1)\end{eqnarray*}
\begin{eqnarray*}
\vec{\nu}_{1,\alpha=2} & = & (0,1,0)\\
\vec{\nu}_{2,\alpha=2} & = & (-1,0,0)\\
\vec{\nu}_{3,\alpha=2} & = & (0,0,1).\end{eqnarray*}
 Denoting $\phi_{i,\alpha}$ as $\phi_{i,1}=\phi_{i}$ and $\phi_{i,2}=\bar{\phi}_{i}$
for the $i$-th spin component with respect to the $\alpha$-th ordering
wave vector, the symmetry operations of the $P4/mmm$ point group
generators are\begin{eqnarray*}
C_{4}[001] & : & \phi_{1}\leftrightarrow\bar{\phi}_{1},\,\phi_{2}\leftrightarrow\bar{\phi}_{2},\,\phi_{3}\longrightarrow\bar{\phi}_{3}\longrightarrow-\phi_{3},\\
C_{2}[100] & : & \phi_{1}\longrightarrow-\phi_{1},\,\phi_{2}\longrightarrow\phi_{2},\,\phi_{3}\longrightarrow\phi_{3},\\
 &  & \bar{\phi}_{1}\longrightarrow-\bar{\phi}_{1},\,\bar{\phi}_{2}\longrightarrow\bar{\phi}_{2},\,\bar{\phi}_{3}\longrightarrow-\bar{\phi}_{3},\\
i & : & \phi_{j}\longrightarrow-\phi_{j},\,\bar{\phi}_{j}\longrightarrow-\bar{\phi}_{j}.\end{eqnarray*}
 We see that once a spin points along one of the principal axes (labeled
by the $i$ index) the application of the crystallographic point group
symmetry operations will ``trap'' it on the same direction. This
results in 3 irreducible representations of size $N=m=2$.

\end{document}